\documentclass[12pt]{article}
\pdfoutput=1
\usepackage{latexsym, graphicx,cite}
\usepackage{amsmath}
\usepackage{amssymb}
\usepackage{amsthm}
\usepackage{graphicx}
\usepackage{subcaption}
\allowdisplaybreaks

\usepackage[top = 1. in,bottom = 1. in, left = 1 in, right=1 in]{geometry}

\newcommand{\arXiv}[1]{\href{http://www.arXiv.org/abs/#1}{arXiv:#1}}
\usepackage[colorlinks=true, linkcolor=blue, bookmarks=true]{hyperref}

\makeatletter
\renewcommand\section{\@startsection {section}{1}{\z@}%
	{-3.5ex \@plus -1ex \@minus -.2ex}
	{2.3ex \@plus.2ex}%
	{\normalfont\large\bfseries}}
\renewcommand\subsection{\@startsection{subsection}{2}{\z@}%
	{-3.25ex\@plus -1ex \@minus -.2ex}%
	{1.5ex \@plus .2ex}%
	{\normalfont\bfseries}}
\makeatother

\newcommand{\beq}{\begin{equation}}
\newcommand{\eeq}{\end{equation}}
\newcommand{\ber}{\begin{array}}
	\newcommand{\eer}{\end{array}}

\newcommand{\de}{\delta}

\newcommand{\eps}{\varepsilon}

\newcommand{\om}{\omega}

\newcommand{\ena}{\end{eqnarray}}
\newcommand{\beqa}{\begin{eqnarray}}
\newcommand{\eeqa}{\end{eqnarray}}
\newcommand{\bea}{\begin{eqnarray}}
\newcommand{\eea}{\end{eqnarray}}

\theoremstyle{remark}

\begin{document}
	\begin{titlepage}
		\begin{flushright}
			\phantom{arXiv:yymm.nnnn}
		\end{flushright}
		\begin{center}
			{\LARGE\bf  Energy returns in global AdS$_4$}\\
			\vskip 10mm
			{\large Anxo Biasi,$^{a}$ Ben Craps$^{\hspace{0.5mm}b}$ and Oleg Evnin$^{c,b}$}
			\vskip 7mm
			{\em $^a$ Departamento de F\'\i sica de Part\'\i culas, Universidade de Santiago de Compostela
				and\\
				Instituto Galego de F\'\i sica de Altas Enerx\'\i as (IGFAE), \\
				E-15782 Santiago de Compostela, Spain}
			\vskip 3mm
			{\em $^b$ Theoretische Natuurkunde, Vrije Universiteit Brussel (VUB) and\\
				The International Solvay Institutes,
				Pleinlaan 2, B-1050 Brussels, Belgium}
			\vskip 3mm
			{\em $^c$ Department of Physics, Faculty of Science, Chulalongkorn University,\\
				Thanon Phayathai, Pathumwan, Bangkok 10330, Thailand}
			\vskip 7mm
			{\small\noindent {\tt anxo.biasi@gmail.com, Ben.Craps@vub.be, oleg.evnin@gmail.com}}
		\end{center}
		\vspace{10mm}
		\begin{center}
			{\bf ABSTRACT}\vspace{3mm}
		\end{center}
		
		Recent studies of the weakly nonlinear dynamics of probe fields in global AdS$_4$ (and of the nonrelativistic limit of AdS resulting in the Gross-Pitaevskii equation) have revealed a number of cases with exact dynamical returns for two-mode initial data. In this paper, we address the question whether similar exact returns are present in the weakly nonlinear dynamics of gravitationally backreacting perturbations in global AdS$_4$. In the literature, approximate returns were first pointed out numerically and with limited precision. We first provide a thorough numerical analysis and discover returns that are so accurate that it would be tantalizing to sign off the small imperfections as an artifact of numerics. To clarify the situation, we introduce a systematic analytic approach by focusing on solutions with spectra localized around one of the two lowest modes. This allows us to demonstrate that in the gravitational case the returns are not exact. Furthermore, our analysis predicts and explains specific integer numbers of direct-reverse cascade sequences that result in particularly accurate energy returns (elaborate hierarchies of more and less precise returns arise if one waits for appropriate longer multiple periods in this manner). In addition, we explain, at least in this regime, the ubiquitous appearance of direct-reverse cascades in the weakly nonlinear dynamics of AdS-like systems. 
		
		\vfill
		
	\end{titlepage}
	
	
	\section{Introduction}
	
	Investigations of weakly nonlinear dynamics of global Anti-de Sitter (AdS) spacetime, initiated by the numerical evidence for its nonlinear instability presented in \cite{BR}, have resulted in a number of surprises. This includes evidence for turbulent cascades leading to black hole formation starting from small initial data \cite{BR}, as well as non-collapsing initial configurations \cite{MR,BS,BScomment,DFLY,Kim,DF}, and a number of intriguing results within the resonant approximation of the AdS dynamics accurately describing slow weakly nonlinear transfer of energy between the linearized modes \cite{FPU,CEV1,CEV2}: selection rules prohibiting certain interaction channels between the linearized modes \cite{CEV1,CEV2,Yang,EN}, extra conservation laws \cite{BKS,CEV2,dual}, dual energy cascades \cite{dual}, and strong numerical evidence for turbulence within the resonant system \cite{BMR} (for a review, see \cite{rev2}). While the bulk of these considerations have focused on the case of spherically symmetric perturbations of gravity-scalar field systems, extensions to pure gravity within squashed sphere ans\"atze \cite{ads5} and more general perturbations outside spherical symmetry (starting with \cite{DHS}) have also appeared.
	
	Among the many surprises of the sort mentioned above is the observation of Fermi-Pasta-Ulam (FPU)-like returns of energy configurations in \cite{FPU}. Conventionally, the Fermi-Pasta-Ulam paradox refers to surprisingly close returns of the energy distribution between linearized modes to its initial configuration observed in the pioneering numerical study \cite{FPU_orig} of weakly nonlinear oscillator chains (for a review, see \cite{FPU_rev}). In the context of weakly nonlinear dynamics of AdS, the energy transfer between the linearized modes is accurately captured by the approximate resonant system, also called the Two-Time Framework (TTF) in \cite{FPU} and the renormalization flow equation or the time-averaged system in \cite{CEV1,CEV2}. One remarkable observation of \cite{FPU} is that, for initial data with all of the energy in the two lowest modes, this resonant system shows a close return (within a few percent) of the energy distribution to the initial configuration after a few direct and reverse cascades of energy to shorter wavelength modes and back. This is reminiscent of the Fermi-Pasta-Ulam phenomenon (though we emphasize that there are also some significant differences: from the standpoint of the effective resonant system, the dynamics is strongly nonlinear, as the weak nonlinearity parameter has been completely scaled out, while the conventional Fermi-Pasta-Ulam paradox is essentially weakly nonlinear).
	
	A number of developments that have occurred since the publication of \cite{FPU} call us to re-examine the issue of energy returns in global AdS$_4$. The complexity of the gravitational dynamics in AdS, even when treated within the resonant approximation, has encouraged examination in \cite{CF,BHP,BEL,BHP2} of a number of related systems for nonlinear probe fields in AdS, without gravitational interactions. These systems, while possessing much simpler nonlinearities, share many structural features of the gravitational problem, including the patterns in the effective resonant system. It turned out that such probe field resonant systems derived in \cite{CF,BHP,BEL}  possess remarkable analytic features, including \emph{exact} energy returns for two-mode initial data. We also mention the closely related problems arising from nonlinear Schr\"odinger equations in isotropic harmonic traps  \cite{GHT,GT,BMP,BBCE,GGT,BBCE2,AO,SNH}. These systems likewise display perfect energy returns, and their close relation to the AdS systems is explained by the nonlinear Schr\"odinger equation in a harmonic trap emerging as a nonrelativistic limit of nonlinear wave equations in AdS \cite{BEL}.
	
	In view of the exact returns for two-mode initial data which can be analytically demonstrated for a number of systems closely related to weakly nonlinear gravitational dynamics in AdS, a question naturally comes up: could it be that the returns are in fact perfect and the imperfections observed in the simulations of \cite{FPU} are a numerical artifact? Indeed, the simulations of \cite{FPU} were performed at a modest numerical precision, which typically involved truncating the system to 30 or 47 lowest modes out of the infinite tower of modes. While a few subsequent studies, including \cite{BMR} and \cite{islands}, were performed at higher numerical precision, they did not target the question of energy returns, and made no comments on this issue. We therefore revisit the problem, and examine the returns of two-mode initial data in global AdS$_4$ in hope of elucidating the question whether such returns are exact or approximate.
	
	The course of our investigation involves both numerical and analytic parts. After reviewing in section~\ref{sec:preliminaries}  the basic setup of the resonant approximation to nonlinear gravitational dynamics in global AdS$_4$, we dedicate section~\ref{sec:numerics} to a numerical implementation of this resonant dynamical system at a much higher precision than in \cite{FPU}, specifically targeting the question of dynamical returns for two-mode initial data. In this way, we discover strikingly accurate returns that are visually indistinguishable from the perfect returns for related systems observed in \cite{CF,BBCE,BEL,BBCE2,AO,SNH}. Nonetheless, the tiny deviation from perfect returns does not appear to go away completely with increased numerical precision. We highlight the subtleties of numerical simulations of the type of problems we are considering, and are forced to look for analytic clarification of the paradoxical situation we observe. 
	
	With this goal in mind we turn, in section~\ref{sec:analytics}, to regimes in which our two-mode initial data are close to one-mode initial data. In this situation, the problem becomes analytically tractable and one can see that the returns are inexact. This should be contrasted with the scenario of \cite{CF,BBCE,BEL,BBCE2,AO,SNH}, where, for related systems, generic two-mode initial data display exact returning behaviors. We thus confirm the FPU-like nature of the energy returns demonstrated by the weakly nonlinear gravitational dynamics in global AdS$_4$, which turn out to display a level of return accuracy much more striking than what has been seen in the past literature. 
	
	Furthermore, our analysis of solutions dominated by one of the two lowest modes predicts and explains specific integer numbers of direct-reverse cascade sequences that result in particularly accurate energy returns (elaborate hierarchies of more and less precise returns arise if one waits for appropriate longer multiple periods in this manner). We provide robust predictions for the numbers of cascades that lead to enhanced returns, even for initial data with moderate ratios of the initial mode energies. In addition, our analytic work explains (at least in the regime of two-mode initial data dominated by one of the modes) the ubiquitous appearance of direct-reverse energy cascade sequences in the weakly nonlinear dynamics of AdS-like systems.  We conclude with a discussion of possible implications in section~\ref{sec:discussion}.
	
	\section{Preliminaries}\label{sec:preliminaries}
	
	We start with a very brief review of the basic setup of the spherically symmetric AdS-scalar field system and the resonant approximation to its weakly nonlinear dynamics. More details can be found in \cite{CEV1,CEV2,rev2}.
	
	One considers Einstein's gravity with a negative cosmological constant 
	\beq
	\Lambda=-\frac{d(d-1)}{2}
	\eeq 
	in $d$ spatial dimensions, coupled to  a free massless scalar field. The equations of motion are
	\begin{equation}\label{Einstein}
	R_{\mu\nu}-\frac{1}{2}g_{\mu\nu}R+\Lambda g_{\mu\nu}-8\pi G\left(\partial_{\mu}\phi\partial_{\nu}\phi-\frac{1}{2}g_{\mu\nu}(\partial\phi)^{2}\right)=0
	\end{equation}
	and
	\begin{equation}\label{scalar}
	\frac{1}{\sqrt{-g}}\partial_{\mu}\left(\sqrt{-g}g^{\mu\nu}\partial_{\nu}\phi\right)=0.
	\end{equation}
	One can consistently truncate to spherically symmetric configurations, corresponding to the metric ansatz
	\begin{equation}\label{eqn:MetricAnsatz}
	ds^{2}=\frac{1}{\cos^{2}x}\left(\frac{dx^{2}}{A}-Ae^{-2\delta}dt^{2}+\sin^{2}x\,d\Omega_{d-1}^{2}\right),
	\end{equation}
	where $A(x,t)$, $\delta(x,t)$ and $\phi(x,t)$ only depend on the time coordinate $t$ and the radial coordinate $x$, which is defined on the interval $[0,\pi/2)$. We shall set  $8\pi G=d-1$.
	
	Following \cite{BR}, we introduce $\Phi\equiv\phi'$ and $\Pi\equiv A^{-1}e^{\delta}\dot{\phi}$ (where dots and primes denote the $t$- and $x$-derivatives, respectively), and also the following two predefined functions
	\begin{equation}
	\mu(x)\equiv(\tan x)^{d-1}
	\qquad\text{and}\qquad
	\nu(x)\equiv\frac{(d-1)}{\mu'(x)}=\frac{\sin x\cos x}{(\tan x)^{d-1}}.
	\label{munu}
	\end{equation}
	The equations of motion are then written as
	\begin{subequations}
		\label{eqn:EOM}
		\begin{align}
		\dot{\Phi}&=\left(Ae^{-\delta}\Pi\right)',
		&\dot{\Pi}&=\frac{1}{\mu}\left(\mu Ae^{-\delta}\Phi\right)', \\
		A'&=\frac{\nu'}{\nu}\left(A-1\right)-\mu\nu\left(\Phi^{2}+\Pi^{2}\right)A,
		&\delta'&=-\mu\nu\left(\Phi^{2}+\Pi^{2}\right), \label{eqn:EOMConstraint}
		\end{align}
		\begin{equation}
		\dot{A}=-2\mu\nu A^{2}e^{-\delta}\Phi\Pi.
		\end{equation}
	\end{subequations}
	Static solutions of these equations are the AdS-Schwarzschild black holes $A(x,t)=1-M\nu(x)$, $\delta(x,t)=0$ and $\phi(x,t)=0$, while unperturbed AdS$_{d+1}$ corresponds to $A=1$, $\delta=\phi=0$.
	
	Note that because of the spherical symmetry imposed by our ansatz (\ref{eqn:MetricAnsatz}), the metric has no propagating degrees of freedom. On each given time slice, the constraint equations (\ref{eqn:EOMConstraint}) can be integrated to express the metric components in terms of the matter distribution given by $\phi(x,t)$ at the same moment of time. Effectively, $A$ and $\de$ can be completely integrated out leaving a nonlinear equation for $\phi$ of the form
	\beq
	\Box_{\mbox{\scriptsize AdS}} \,\phi = S[\phi],
	\label{nlinwave}
	\eeq
	where $\Box_{\mbox{\scriptsize AdS}}$ is the Laplacian in (a non-dynamical) AdS$_{d+1}$ and $S$ symbolically denotes all the nonlinear terms (which are local in time but nonlocal in space) arising from integrating out $A$ and $\de$. One can effectively construct $S$ as an expansion in powers of $\phi$, and only the cubic part is significant for our present discussion, since that is what affects the specific weakly nonlinear regime we focus on, see \cite{CEV2}. (One might find it strange that we talk about gravitational dynamics in AdS, while under the assumption of spherical symmetry the metric is non-dynamical and one ends up with a nonlinear wave equation for a scalar field in a fixed AdS background. Nonetheless, there are closely related constructions, which are purely gravitational, without any matter, and utilize a `squashed' generalization of our ansatz, see \cite{ads5}. Such extensions effectively result in nonlinear wave equations very similar to the one we have for the scalar field, but now satisfied by the metric components in the absence of any matter.)
	
	The problem of weakly nonlinear gravitational dynamics of the AdS-scalar field system under the assumption of spherical symmetry is thus reduced to a complicated nonlinear wave equation in a fixed AdS background. This highlights the relation between gravitational stability of AdS and simpler nonlinear wave equations in AdS that have been considered in the literature as toy models, for example, the $\lambda \phi^4$ theory in nondynamical AdS, see \cite{BKS,CF,BEL}. Our aim shall be to develop an effective treatment of this dynamics for small fields $\phi$ of order $\eps$ on long time scales of order $1/\eps^2$. This is the regime in which interesting phenomenology, including black hole formation (`turbulent instability'), has been observed in numerical experiments starting with \cite{BR}.
	
	Before proceeding with weakly nonlinear analysis, one must thoroughly understand the linearized problem, i.e. the equation
	\beq
	\Box_{\mbox{\scriptsize AdS}} \,\phi =0,
	\label{linwave}
	\eeq
	which under the assumption of spherical symmetry takes the form
	\begin{equation}\label{eqn:phi1}
	\ddot{\phi}_{1}+\hat{L}\phi_{1}=0
	\qquad\text{with}\qquad
	\hat{L}\equiv-\frac{1}{\mu(x)}\partial_{x}\left(\mu(x)\partial_{x}\right).
	\end{equation}
	The eigenvalues for the operator $\hat{L}$ are $\omega_{n}^{2}$, with
	\begin{equation}\label{spectr}
	\omega_n=d+2n,\ \ \  n=0,1,... ,
	\end{equation}
	and  the eigenfunctions are
	\begin{equation}
	e_{n}(x)=k_{n}\cos^{\,d} x\,P_{n}^{\left(\frac{d}{2}-1,\frac{d}{2}\right)}\left(\cos 2x\right)
	\qquad\text{with}\qquad
	k_{n}=\frac{2\sqrt{n!(n+d-1)!}}{\Gamma\left(n+\frac{d}{2}\right)}.
	\label{modes}
	\end{equation}
	Here, $P_{n}^{(a,b)}(x)$ are Jacobi polynomials of degree $n$.
	
	A remarkable feature, intimately linked to the isometries of AdS$_{d+1}$ forming the conformal group $SO(d,2)$, is that all solutions oscillate with integer frequencies $ \omega_n$. This, in fact, extends outside spherical symmetry, and the most general solution of (\ref{linwave}) is time-periodic with period $2\pi$. (The entire tower of eigenmodes of (\ref{linwave}) fills an infinite-dimensional representation of $SO(d,2)$ -- and it is also related by a simple transformation to wavefunctions of a superintegrable quantum-mechanical system known as the Higgs oscillator \cite{higgs,KGize}.) 
	
	The fully resonant spectrum (\ref{spectr}) is highly atypical, and it gives arbitrarily small nonlinearities an opportunity to have large effects, since the impact of resonant interactions between the modes tends to accumulate over time. In naive perturbation theory in terms of power series in $\eps$, this feature is reflected in a breakdown of perturbative expansions on time scales of order $1/\eps^2$, as noted already in \cite{BR}. More specifically, if one tries to develop solutions of (\ref{nlinwave}) as solutions of (\ref{linwave}) of order $\eps$ plus corrections of higher orders in $\eps$, already at order $\eps^3$ one ends up with `secular' terms growing as $\eps^3 t$, which overpower the leading (linear) term on time scales greater than $1/\eps^2$. The origin of such terms can be directly traced back to the presence of resonances in the linearized spectrum. 
	
	The discussion above shows that naive perturbative expansions are not a valid way to approximate the weakly nonlinear dynamics of (\ref{nlinwave}) on the relevant timescales. A particularly viable alternative approach is time-averaging, introduced for studies of AdS systems in \cite{FPU} and developed analytically in \cite{CEV1,CEV2} (various names were used for it). One starts by expanding solutions of (\ref{nlinwave}) in terms of linearized eigenmodes (\ref{modes}) as
	\begin{align}
	&\phi(x,t)=\eps\sum_{n=0}^{\infty} \left(\alpha_n(t)e^{-i\omega_n t}+\bar\alpha_n(t)e^{i\omega_n t}\right) e_n(x),\label{phiexp}\\ 
	&\dot\phi(x,t)=-i\eps\sum_{n=0}^{\infty} \omega_n\left(\alpha_n(t)e^{-i\omega_n t}-\bar\alpha_n(t)e^{i\omega_n t}\right) e_n(x),\nonumber
	\end{align}
	and equivalently rewriting (\ref{nlinwave}) as a system of equations for $\alpha_n$. This system of equations for $\alpha_n$ is of course completely equivalent to (\ref{nlinwave}), but it matches what is known as the `periodic standard form' in mathematical literature, and the time-averaging procedure may be applied to simplify this equation in a way that does not affect its accuracy on timescales of order $1/\eps^2$. More specifically, substituting (\ref{phiexp}) in (\ref{nlinwave}) and projecting on $e_n$, one expresses $i\dot\alpha_n$ as a cubic combination of $\alpha_k$ and $\bar\alpha_k$. Most of the terms in this cubic combination come with oscillatory factors inherited from the explicit time dependences in (\ref{phiexp}). Time-averaging (backed by precise mathematical theorems) discards all oscillatory terms, retaining only the {\it resonant} terms in which the oscillatory factors cancel. This is guaranteed not to affect accuracy on time scales of order $1/\eps^2$ and results in a {\it resonant system} of the form
	\beq
	i\omega_n\dot\alpha_n=\hspace{-5mm}\sum_{\om_n+\om_m=\om_k+\om_l}\hspace{-5mm} C_{nmkl}\,\bar\alpha_m\alpha_k\alpha_l,
	\label{ressyst}
	\eeq
	where, from now on, the dot will stand for a derivative with respect to the \emph{slow time} $\eps^2 t$, which we shall call simply $t$ from now on.
	A few remarks are in order:
	\begin{itemize}
		\item The dependence on the small nonlinearity parameter has been completely scaled out of (\ref{ressyst}) with the introduction of the slow time, and there are no small parameters left. In fact, (\ref{ressyst}) enjoys a {\it scaling symmetry}: if $\alpha_n(t)$ is a solution, so is $\lambda\alpha_n(\lambda^2t)$, for any $\lambda$.
		\item The nonlinear physics of the problem is completely encoded in the {\it interaction coefficients} $C_{nmkl}$ which are a set of numbers expressed through the mode functions (\ref{modes}) and their spatial derivatives, and depending on the specific form of nonlinearity in (\ref{nlinwave}). For the gravitational system we are considering here, the explicit expressions for $C$ are very complicated and can be found in \cite{CEV2} (where these coefficients are denoted as $T$, $R$ and $S$, depending on the number of coincident index values). In the Appendix, we extend them to the case of a massive scalar field, which will be discussed in section~\ref{sec:discussion}. 
		\item The frequencies $\omega_n$ can be absorbed in $C$ by redefining $\alpha_n$, but it is often convenient to keep them explicit. Note that $\om_n+\om_m=\om_k+\om_l$ is equivalent to simply $n+m=k+l$.
		\item A number of simpler related problems (nonlinear wave equations in non-dynamical AdS, variations of the Gross-Pitaevskii equation for Bose-Einstein condensates in harmonic traps) lead to resonant systems that only differ from (\ref{ressyst}) by specific values of the numerical coefficients. Some of these systems demonstrate perfect energy returns in the evolution of two-mode initial data \cite{CF,BEL,BBCE,BBCE2}.
		\item Purely from resonance analysis, (\ref{ressyst}) could have extra terms (for example, with three $\alpha$'s and no $\bar\alpha$'s), but such terms can be shown to vanish specifically in AdS due to special selection rules \cite{CEV1,CEV2,Yang,EN}. Correspondingly, there is an emergent $U(1)$ symmetry rotating the phases of $\alpha_n$ by a common shift, which is manifest in (\ref{ressyst}) but absent in (\ref{nlinwave}).
		\item In immediate relation to the extra $U(1)$ symmetry resulting from selection rules, there is an extra conservation law \cite{BKS,CEV2,dual} for the associated `particle number'
		\beq
		N=\sum\om_n|\alpha_n|^2
		\label{consN}
		\eeq
		in addition to the total `linearized energy' 
		\beq
		E=\sum\om_n^2|\alpha_n|^2,
		\label{consE}
		\eeq
		generically conserved by resonant systems. They are associated with the symmetry transformations
		\beq
		\alpha_n \rightarrow e^{i\varphi}\alpha_n,\qquad \alpha_n \rightarrow e^{i \om_n\theta}\alpha_n,
		\label{symtransNE}
		\eeq
respectively.
	\end{itemize}
	
	Our main objective is to examine solutions to (\ref{ressyst}) starting with two-mode initial data
	\beq
	\alpha_0(0),\alpha_1(0)\ne 0,\qquad \alpha_{n\ge 2}(0)=0.
	\label{twomode}
	\eeq
	In particular, the energy distribution between the modes may be quantified by 
	\beq
	E_n(t)=\om_n^2|\alpha_n(t)|^2,
	\eeq
	and we may scan the evolution of two-mode initial data for return moments when the deviation from the initial energy distribution is small.
	
	Due to the scaling symmetry of (\ref{ressyst}) we can transform any initial data (\ref{twomode}) to satisfy 
	\beq
	E_0(0)+E_1(0)=1.
	\label{energynorm}
	\eeq 
	Thereafter, it is sufficient to study the quantity
	\beq
	\Delta(t)=1-E_0(t)-E_1(t).
	\label{deltadef}
	\eeq
	If $\Delta(t)$ is 0, joint conservation of (\ref{consN}) and (\ref{consE}) guarantees that $E_0(t)=E_0(0)$ and  $E_1(t)=E_1(0)$, i.e., we have found a perfect return, while $0<\Delta(t)\ll 1$ signifies an accurate but imperfect return. As we have already remarked, for a number of related resonant systems solved in the literature, $\Delta(t)$ periodically drops to 0 for any two-mode initial data of the form (\ref{twomode}), signalling exact returns \cite{CF,BEL,BBCE,BBCE2}. In our present work, we shall look closely into similar return phenomena for the gravitational resonant system (\ref{ressyst}), which is only possible numerically.
	
	\section{Numerics}\label{sec:numerics}
	
	Before describing the results of our own numerical experiments, we briefly summarize what has been done in this area before. Already in \cite{FPU}, one of the themes was the observation of accurate but imperfect returns of energy to the initial configuration for two-mode initial data in the resonant system for gravitational AdS$_4$ perturbations (there referred to as the Two-Time-Framework approximation). These numerical studies were performed at the resolution available at that time, which amounted to truncating the resonant system to the lowest 30 or 47 modes, and returns with precision of a few percent were observed. The evolution was seen to proceed in a sequence of direct and reverse energy cascades, and the third reverse cascade led to a more accurate return to the initial energy distribution than the first two. Subsequently, a few developments occurred, including the derivation of analytic expressions for the interaction coefficients of the AdS resonant systems in terms of the AdS mode functions in \cite{CEV1,CEV2} and an algorithm to convert these expressions into explicit (but very complicated) functions of the mode numbers in \cite{islands}. While this has allowed simulations of the resonant dynamics with much higher precision, the question of energy returns has not been properly readdressed in the literature following these developments.
	
	We reiterate our reasons to revisit the question of energy returns in AdS$_4$. Our current perspective is quite different from the predominant views at the time \cite{FPU} was written. Indeed, in the years that have passed, a number of examples have emerged, where resonant systems closely related to the one studied in \cite{FPU} display \emph{exact}, rather than approximate, energy returns. These include nonlinear probe fields in AdS \cite{CF, BEL} as well as nonlinear Schr\"odinger equations in harmonic potentials that arise from AdS systems in a nonrelativistic limit \cite{BBCE,BBCE2,SNH}. This puts the results of \cite{FPU} in a different light and calls for their re-examination. Note that the limited precision of \cite{FPU} (mode number cut-off at a few dozens modes) makes it impossible to distinguish exact and approximate returns. In particular, in application to the systems of \cite{CF, BEL, BBCE,BBCE2,SNH}, where the returns are \emph{analytically} known to be exact, such simulations would have indicated approximate returns (the imperfection in this case being a pure artifact of truncating the infinite-dimensional system).
	
	\begin{figure}[t]
		\begin{subfigure}[b]{.51\linewidth}
			\includegraphics[scale=0.54]{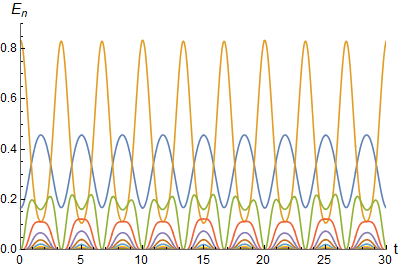}
			\caption{$\text{AdS}_4$}\label{fig:AdS_9_oscillations}
		\end{subfigure} 
		\begin{subfigure}[b]{.4\linewidth}
			\hspace{-1cm}
			\includegraphics[scale=0.54]{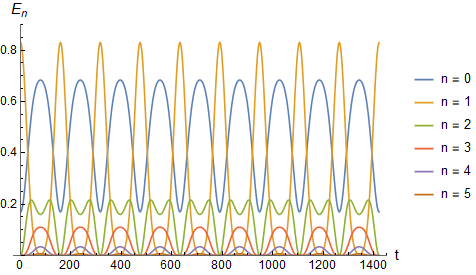}
			\caption{LLL}\label{fig:LLL_9_oscillations}
		\end{subfigure}
		\caption{Nearly periodic dynamics of the spectrum for the gravitational AdS$_4$ resonant system vs. exactly periodic dynamics for the LLL equation. The tiny deviations from exact periodicity in (a) are not visually discernible. The initial condition is (\ref{twomode}, \ref{energynorm})  with $E_0 =0.17$ for both plots.}
		\label{fig3periods}
	\end{figure}
	\begin{figure}[t]
		\begin{subfigure}[b]{.45\linewidth}
			\includegraphics[scale=0.5]{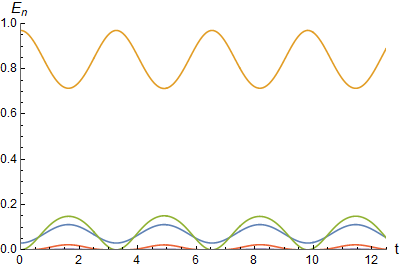}
			\caption{$E_0 = 0.03$}\label{fig:two_mode_evol_sequence_A}
		\end{subfigure}
		\begin{subfigure}[b]{.45\linewidth}
			\includegraphics[scale=0.5]{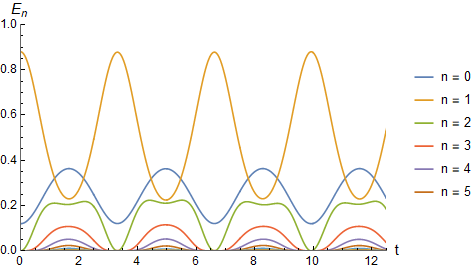}
			\caption{$E_0 = 0.12$}\label{fig:two_mode_evol_sequence_B}
		\end{subfigure}  
		
		\begin{subfigure}[b]{.45\linewidth}
			\includegraphics[scale=0.5]{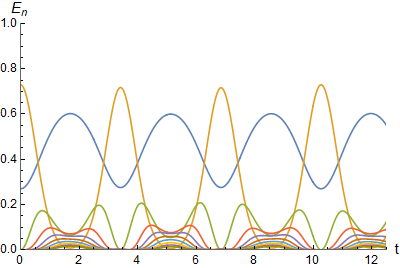}
			\caption{$E_0 = 0.27$}\label{fig:two_mode_evol_sequence_C}
		\end{subfigure}
		\begin{subfigure}[b]{.45\linewidth}
			\includegraphics[scale=0.5]{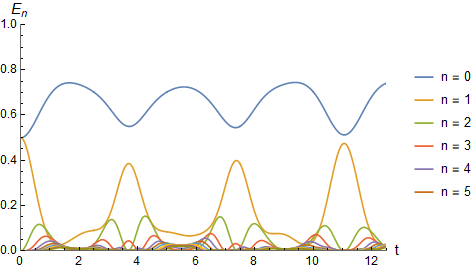}
			\caption{$E_0 = 0.5$}\label{fig:two_mode_evol_sequence_D}
		\end{subfigure}  
		\caption{This sequence of time evolutions of four two-mode initial data (\ref{twomode}, \ref{energynorm}) shows that (a) the energy transfer between modes appears to evolve in a periodic way when the initial condition is close to mode 1 (a), that for larger $E_0$ the evolution is apparently periodic only after three oscillations (b,c), and that for equal-energy initial conditions, the returns are visibly not exact, but still more accurate after three oscillations (d).}
		\label{fig:two_mode_evol_sequence}
	\end{figure}
	To elucidate the issue of energy returns in global AdS$_4$, we have performed simulations with up to 500 modes studying the dependence of the return accuracy on the mode number cut-off as well as on the arithmetic precision. For some initial energy distributions between the two lowest modes, the returns we observe are strikingly accurate. Fig.~\ref{fig:AdS_9_oscillations} provides an illustration in this regard: we study the evolution of two-mode initial data which are fairly generic (not particularly close to one-mode initial data) over a sequence of nine direct-reverse cascades. After each three cascades, we have a return to the initial configuration that is visually indistinguishable from exact, and furthermore the whole pattern periodically repeats after each three oscillations in a way that is visually indistinguishable from exact periodicity. All of this happens in (our high-precision numerical truncation of) the infinite-dimensional nonlinear resonant system (\ref{ressyst}) that has no small parameters! Note that the plot given in fig.~\ref{fig:AdS_9_oscillations} is visually extremely similar to an analogous plot (see fig.~\ref{fig:LLL_9_oscillations}) in a closely related resonant system called the LLL equation, where the energy returns have been analytically proven to be exact \cite{BBCE}. Fig.~\ref{fig:two_mode_evol_sequence} provides four illustrations where we can observe the behavior of the energy flow for different initial energy distributions between the first two modes: fig.~\ref{fig:two_mode_evol_sequence_A} is particularly close to the second mode and the evolution appears periodic after each direct-reverse cascade, while in fig.~\ref{fig:two_mode_evol_sequence_B} and fig.~\ref{fig:two_mode_evol_sequence_C} the apparent periodicity only emerges after three direct-reverse cascades. Finally fig.~\ref{fig:two_mode_evol_sequence_D} describes initial data for which the evolution visibly deviates from perfect periodicity, although it is remarkable that the deviations are still pretty small.

\begin{figure}[t]
                   \begin{subfigure}[b]{.5\linewidth}
		\includegraphics[scale=0.55]{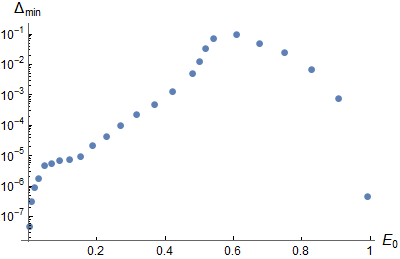}   
                    \caption{}
                   \end{subfigure}\hspace{5mm}
                    \begin{subfigure}[b]{.5\linewidth}
                    \includegraphics[scale=0.55]{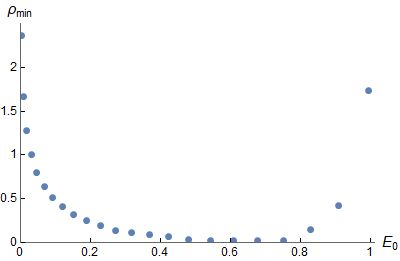}
                   \caption{}
                   \end{subfigure}
		\caption{Energy return precision (a) and the minimal spectral suppression exponent (b) plotted against the initial energy of mode 0 for two-mode initial data (\ref{twomode}, \ref{energynorm}). Small values of the curve in (b) correspond to strong turbulent cascades of energy.}
		\label{initcond}
	\end{figure}

\begin{figure}[t]
		\begin{subfigure}[b]{.51\linewidth}
			\includegraphics[scale=0.54]{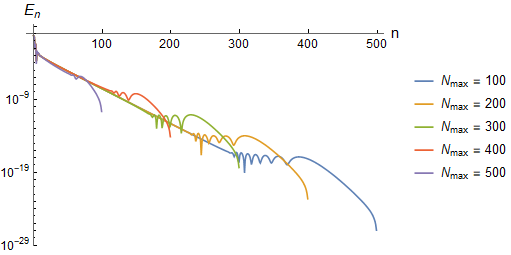}
			\caption{$E_0 = 0.5,\ t = 3.75$}\label{fig:cut_off_effects}
		\end{subfigure} 
		\begin{subfigure}[b]{.4\linewidth}
			\hspace{-1cm}
			\includegraphics[scale=0.54]{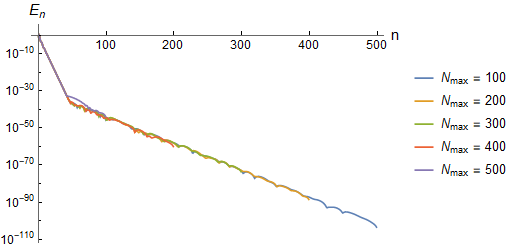}
			\caption{$E_0 = 0.17,\ t = 3.5$}\label{fig:shelf_artifects_effects}
		\end{subfigure}
		\caption{Energy spectrum of the two-mode initial data (\ref{twomode}, \ref{energynorm}) at the bottom of the first reverse cascade for different mode number cut-off. In (a), the direct cascade is strong, in which case the most relevant effects are produced by the cut-off, resulting in spurious oscillatory behaviors in the upper half of the mode range. In (b), the direct cascade is weak, and `shelf' artifacts dominate the imperfections of the spectrum. The `shelf' here refers to the shallower portion of the spectrum starting around mode 50. The true dynamics of the underlying system is expected to be represented by a smooth extension of the steeply downward, approximately straight line in the lower portion of the spectrum, rather than by the `shelf' artifact. Further discussion can be found in the text.}
		\label{fig:numerical_artifacts}
	\end{figure}

	And yet, spectacular as some of these returns are, they are not exact. This by itself is not conclusive, since returns in numerical simulations of truncated systems would not have been exact even if returns in the infinite-dimensional dynamical system were exact. We thus have to quantify the dependence of the accuracy of returns on the initial data, as well as on the mode number cut-off, arithmetic precision and other numerical imperfections. To set the stage, we present in fig.~\ref{initcond} plots showing the dependence of two essential quantities on the initial energy of mode 0 for two-mode initial data. 
	The first quantity (fig~\ref{initcond}a) is the minimal value of $\Delta$ defined by (\ref{deltadef}) over the first three direct-reverse cascades, which quantifies the energy return precision (evidently, in computing the minimum we exclude the initial part of the first direct cascade where $\Delta$ is small simply because the system has not had enough time yet to deviate from the initial conditions). We see that the worst returns are around $E_0\approx 0.6$ at a level slightly better than 10\%, while for most other initial data away from that value the returns have precision better than 1\%. The second quantity we plot (in fig~\ref{initcond}b) measures the strength of the turbulent cascade starting with the given two-mode initial data. The energy spectrum of configurations that undergo regular evolution is suppressed exponentially for large mode numbers, $E_n(t)\sim n^{\gamma(t)}\exp(-\rho(t)n)$ with $\rho(t)>0$, and turbulent singularity formation, in particular, would correspond to $\rho(t)$ hitting zero at a certain time \cite{AS,BJ,BMR}. We quantify the strength of a turbulent cascade by determining how small $\rho(t)$ becomes during the time interval of interest. Therefore, we first fit the logarithm of the spectrum $\log E_n(t)$ to $-\rho(t) n+\gamma(t)\log n$ and then plot the minimal value of $\rho$ attained over the first three oscillations. We see that the energy return imperfections are especially strong in the region where the turbulent cascade is strong as well. A strong turbulent cascade means, in particular, a stronger sensitivity to the mode number cut-off (as one would need to appreciably excite modes above the cut-off to keep the evolution exact, but those modes are excluded from simulations). This makes the problem of interpreting the deviations from exact returns rather subtle. 
	Since we are entering the realm of precision questions, we must categorize and quantify the uncertainties incurred by our numerics. There are of course generic small errors arising from numerical integration of the equations of motion, but we feel that the following three aspects are particularly important, since they are specific to the sequences of direct and reverse cascades of energy we study: 
	
	\noindent 1) {\bf Mode number truncation:} Restricting to a finite number of modes is expected to have little effect if the energy mostly remains locked within a set of low-lying modes. In our case, an energy cascade develops, transferring the energy to higher modes. While the cascade is self-limiting (unlike in the fully turbulent case of \cite{BMR}, where $\rho(t)$ develops a zero), it can be quite strong, as evident from fig.~\ref{initcond}b. What we observe in numerical experiments is that, as the cascade hits the mode number cut-off and then starts receding, a comb-like pattern of spikes in the spectrum forms. This pattern undergoes an evolution of its own and remains visible in future direct-reverse cascade oscillations. As we can observe in fig.~\ref{fig:cut_off_effects}, these comb patterns are pure numerical artifacts, and do not reflect the true dynamics of the infinite-dimensional model, since the parts of the spectrum overrun by the comb artifacts disagree between simulations with different mode number cut-offs. Such artifacts are generically visible for numerical simulation of resonant systems, including those for which exact analytic solutions starting with two-mode initial data are known (and do not display such comb-like features). While the total energy in the comb-like artifacts is very small at the moment of their formation, it is difficult to predict how they affect future direct-reverse cascade sequences, since the system is nonlinear and prone to chaotic divergence of trajectories.
		
	\noindent 2) {\bf Arithmetic precision:} There is a very peculiar way in which direct-reverse cascade sequences amplify numerical errors, including the most basic rounding errors caused by the finiteness of arithmetic precision. Indeed, assume that in the exact solution a particular mode number $k$ oscillates between energies $E_k^{max}$ and $E_k^{min}$. In our scenario of near-perfect returns $E_k^{min}\ll E_k^{max}$, while for perfect returns $E_k^{min}=0$. Since we are integrating the equations numerically and with finite precision, $E_k^{max}$ contains a relative error given by at least the arithmetic precision, for example $10^{-15}$. As the cascade recedes and the energy drops, the {\it absolute error} in $E_k$ cannot decrease. Therefore, when one arrives at the bottom of the reverse cascade, the energy is $E_k^{min}$, but the absolute error is still $10^{-15}E_k^{max}$ (in particular, values of $E_k^{min}$ below $10^{-15}E_k^{max}$ can never be reached, even if the exact solution corresponds to  $E_k^{min}=0$). The relative error is now $10^{-15}E_k^{max}/E_k^{min}$, much greater than the rounding errors themselves. This relative error will be transported upstream in the next direct cascade and will result in a large absolute error $10^{-15}(E_k^{max})^2/E_k^{min}$ at the peak of the cascade. Thus we can see that repeated transport of absolute errors downstream and relative errors upstream in direct-reverse cascade sequences leads to strong amplification of numerical imprecision (and would of course compromise exact returns even if they were present in the underlying dynamical system). In practice, we can see the formation of flat `shelf' artifacts in the high mode number part of our spectrum as the first cascade recedes, as in fig. \ref{fig:shelf_artifects_effects} (these artifacts precisely reflect the inability of the spectum to go below  $10^{-15}E_k^{max}$ in numerical simulations). Such artifacts continue to evolve in future spectrum oscillations. As the near-perfect returns we observe occur after three direct-reverse cascade cycles, assessing the ultimate impact of these imperfections on the return precision is subtle.
	
	\noindent 3) {\bf Interaction coefficients:} Small errors in the interaction coefficients $C$ can produce important errors in the evolution governed by (\ref{ressyst}). This is particularly important when high modes are involved, as computing the interaction coefficients typically requires numerical integrations of oscillatory functions with frequencies that grow with the mode numbers. These integrals have to be evaluated for each resonant quartet of modes satisfying $n+m=j+k$, and the number of such quartets grows like $O(N_{max}^3)$, where $N_{max}$ is the mode number cut-off. In addition, in order to obtain accurate values of the coefficients $C$ when $N_{max}$ is large, one needs to evaluate their expressions on a very dense grid and with increased arithmetic precision, which further increases the burden. To avoid these problems we computed the fully analytic expression for $C$ in AdS$_4$ described in \cite{islands}, which enables us to evaluate $C$ safely for high modes. The remaining burden is then the computational cost of the simulations, which makes it hard to go beyond $N_{max} = 500$.
	
\begin{figure}[t]
		\centering
		\includegraphics[scale=0.6]{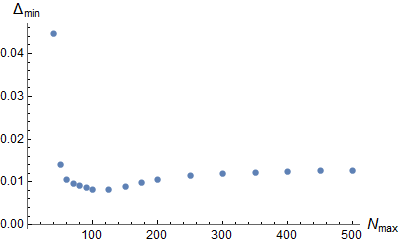}  
		\caption{Energy return precision for two-mode initial data $E_0 = 0.5$ as a function of the mode number cut-off.}
		\label{cutoffdep}
	\end{figure}

	We remark that improving on either 1) or 2) beyond the level of our current simulations would be very demanding in terms of computational costs. While a break-through in our numerical precision does not appear viable at this moment, we have performed some basic comparison of the return precision computed at different values of numerical approximation parameters (as well as a study of artifacts that we have briefly summarized in the passages above). In fig.~\ref{cutoffdep}, we show the dependence of return precision on the number of modes included in our simulations.
	As one can see, the return precision substantially increases as the cut-off is raised from values around 50 (typical of the simulations of \cite{FPU}), but after 200 stabilizes at a small nonzero value. At a naive level, this can seen as an indication that the returns are not exact in the underlying dynamical system (\ref{ressyst}). One has to keep in mind that other factors hindering perfect returns (finite arithmetic precision, possible chaotic enhancement of artifacts over the course of three direct-reverse cascade oscillations) have not been taken (and are very difficult to take) into account.
	
	We conclude that our numerical analysis indicates strikingly accurate, but likely imperfect energy returns for two-mode initial data, though it is not possible as of now to control all possible sources of imperfection. There is, however, one particular aspect that can be further elucidated. Our motivation to take a closer look at energy returns in global AdS$_4$ has largely come from the similarity of the observed dynamics to the known related analytically tractable cases \cite{CF,BEL,BBCE,BBCE2,AO} where the returns are exact. This similarity extends in a few other ways, in particular the ultraviolet parts of the spectra remain approximately exponential for all times (only the slope of their logarithmic plot changes). However, for the systems of \cite{CF,BEL,BBCE,BBCE2,AO}, energy returns for {\it any} two-mode initial data (\ref{twomode}) are exact. This type of dynamics can be excluded for the resonant system (\ref{ressyst}) by inspecting solutions close to mode 0 or mode 1, which are analytically tractable. This will show that the resonant dynamics of AdS$_4$ cannot be strictly in the same class as those of the similar resonant systems of \cite{CF,BEL,BBCE,BBCE2,AO}, which do exhibit perfect energy returns.
	
	\section{Analytics}\label{sec:analytics}
	
	While direct analytic investigations of the resonant system (\ref{ressyst}) are beyond practically imaginable limits, there are special regimes in which this system is analytically tractable. First of all, there are exact single-mode solutions, in which all amplitudes are zero except for one chosen mode \cite{BR}. The vicinities of such single-mode solutions form stability islands \cite{BR,BS,BScomment,FPU,islands}. While it is common to linearize in the vicinity of such single-mode solutions and their generalizations \cite{islands,BHP}, which results in linear systems that are potentially tractable, our approach here will also gain leverage by relying on single-mode solutions, but in a way different from linearization, and much more useful for our purposes.
	
	Instead of linearizing around single-mode solutions, we shall assume that all other modes are exponentially suppressed in proportion to their distance from  the dominant mode. We shall see that, in the limit when this exponential suppression becomes strong, the equations dramatically simplify. Such techniques have been applied in \cite{CEJV} to the analysis of stationary solutions of (\ref{ressyst}) and related resonant systems. While the equations are still nonlinear (unlike in the linearization approach mentioned above), they can be solved iteratively starting with the dominant mode. This structure is very convenient for our purposes, since we are trying to understand whether the energy returns may be exact. If the returns are exact, they are exact for all modes, while a failure of exact returns shall be seen in the approach we are adopting in a finite number of steps, since the discovery of any given non-returning mode guarantees that the returns in the whole system cannot be exact. This procedure has to be repeated two times, for solutions dominated by mode 0 and mode 1, respectively.
	
	
	\subsection{Solutions dominated by mode 0}

	We shall assume that the spectrum is exponentially suppressed as one moves away from the dominant mode 0:
	\begin{equation}\label{SLL0defeq}
	\alpha_{n} = \delta^{n}\, \frac{q_{n}(t)}{\sqrt{\om_n}}, 
	\end{equation}
	with $\delta \ll 1$. Then, after the redefinition $C_{nmkl}^{\rm new} = C_{nmkl}^{\rm old}/\sqrt{\om_n\om_m\om_k\om_l}$, (\ref{ressyst}) becomes
	\begin{equation}\label{resssystde}
	i\dot{q}_{n} = \sum_{m=0}^{\infty}\sum_{k=0}^{n+m}\,\delta^{2m}\, C_{nmk,n+m-k}q_{k}q_{n+m-k}\bar{q}_{m}.
	\end{equation}
	At leading order in $\de$, only the terms with $m=0$ survive:
	\begin{equation}
	i\dot{q}_{n} = \bar{q}_{0}(t)\sum_{k=0}^{n}C_{n0k,n-k}q_{k}q_{n-k}.
	\label{eq:SLL}
	\end{equation}
	While this equation is still nonlinear, it offers a tremendous simplification over (\ref{ressyst}), since it can be solved recursively: once solutions for the modes up to $q_n$ have been constructed, finding $q_{n+1}$ amounts to solving a single linear ODE. We display this structure by writing out the equations for the first few modes:  
	\begin{eqnarray}
	i \dot{q}_{0} & = & C_{0000}|q_{0}|^{2}q_{0},\label{eq:SLL0_eq_0_exp} \\
	i \dot{q}_{1} & = & 2C_{1010}|q_{0}|^{2}q_{1}, \label{eq:SLL0_eq_1_exp}\\
	i \dot{q}_{2} & = & 2 C_{2020}|q_{0}|^{2}q_{2} + C_{2011}\bar{q}_0q_{1}^2,  \label{eq:SLL0_eq_2_exp}\\
	i \dot{q}_{3} & = & 2 C_{3030}|q_{0}|^{2}q_{3} + 2C_{3021}\bar{q}_0q_1q_{2}  , \label{eq:SLL0_eq_3_exp}\\
	i \dot{q}_{4} & = & 2 C_{4040}|q_{0}|^{2}q_{4} + 2C_{4031}\bar{q}_0q_{1}q_{3}   + C_{4022}\bar{q}_0q_{2}^2. \label{eq:SLL0_eq_4_exp}
	\end{eqnarray}
	 By applying the  scaling symmetry of (\ref{ressyst}), one can always set $|q_0|^2$ to 1. In order to elucidate whether energy returns of the two-mode initial data of the form (\ref{twomode}) are exact, we solve (\ref{eq:SLL}) with the initial conditions $q_0(0)=1$, $q_1(0)=1$, $q_{n\ge 2}=0$. (The phases of $q_0(0)$ and $q_1(0)$ can be adjusted as necessary by the symmetry transformations (\ref{symtransNE}), while the magnitude of $q_1(0)$ can be set to 1 by defining $\delta=|\alpha_1(0)|$.) The solutions for $q_0$ and $q_1$ are
	\begin{equation}
	q_{0}(t) = e^{ - i C_{0000}t}, \quad  \quad q_{1}(t) = e^{- 2i C_{1010}t}.
	\end{equation} 
	Note that the energies in these first two leading modes are time-independent, which implies that close to mode 0 the cascade to higher energies is necessarily very weak (the situation will be more interesting for solutions dominated by mode 1).
	For higher $q_n$, one then obtains
	\begin{equation}
	i\dot q_n - 2C_{n0n0} q_n = \bar q_0\sum_{k=1}^{n-1} C_{n0k,n-k} q_k q_{n-k}.
	\label{eq:q_n_mode_0_eq}
	\end{equation}
	The structure of solutions is easily understood recursively: the right-hand side consists of terms oscillating with frequencies given by linear combinations (with integer coefficients) of $C_{0000}$, $C_{1010}$, ..., $C_{n-1,0,n-1,0}$. Hence, $q_n$ will consist of terms that oscillate\footnote{There is a possible caveat in relation to this reasoning in the sense that, if a resonance occurs between the oscillatory terms on the right-hand side of (\ref{eq:q_n_mode_0_eq}) and the frequency $2C_{n0n0}$ on the left-hand side, growing (rather than purely oscillatory) terms will be produced in the solution for $q_n$. However, our explicit computations at a number of low-lying levels, and
preliminary analytic considerations at general $n$ suggest that such resonances do
not in fact occur.} with frequencies given by linear combinations (with integer coefficients) of $C_{0000}$, $C_{1010}$, ..., $C_{n0n0}$. It is only if {\em all} $\,\,q_n$ constructed in this manner (for two-mode initial conditions) oscillate with a common period that one gets exact energy returns.
	
	For the resonant system (\ref{ressyst}), one can straightforwardly proceed with this algorithm, after having computed a few low-lying interaction coefficients from the complicated formulas given in \cite{CEV2}. The result for $q_2(t)$, which is
	\begin{equation}
	q_2(t) = \frac{- 2iC_{2011}}{C_{0000} - 4 C_{1010} + 2C_{2020}}\,\sin\frac{(C_{0000} - 4C_{1010} + 2C_{2020})t}{2} \,e^{i(C_{0000} - 4C_{1010} - 2C_{2020})\frac{t}{2}},
	\end{equation}
	shows that $|q_2(t)|^2$ is periodic with period $T_2 = \frac{10}{123}\pi^2$. Thus, energy returns are perfect at this order, but the period $T_2$ is broken at the next order given by
	\begin{equation}
	q_3(t) = \frac{2C_{2011}C_{3021}}{\lambda_1\lambda_2(\lambda_2-\lambda_1)}\left((\lambda_2-\lambda_1) - \lambda_2 e^{-i \lambda_1 t} + \lambda_1 e^{- i \lambda_2 t}\right) e^{-i 2C_{3030} t},\label{SLL0_q3}
	\end{equation}
	where 
	\begin{align}
	&\lambda_1 = - 2C_{0000} + 6C_{1010} - 2 C_{3030},\\
	&\lambda_2 = - C_{0000} + 2 C_{1010} + 2C_{2020} - 2C_{3030}.
	\end{align} 
	In our particular case their specific values are: $\lambda_1 = \frac{3057}{70\pi}$, $\lambda_2 = \frac{267}{14\pi}$ meaning that at $|q_3(t)|^2$ the energy flow has a periodicity of $T_3 = \frac{140 \pi^2}{3}$, so that $T_3$ is exactly $574$ times greater than $T_2$.  Inspecting the next step of our iterative solution, we conclude that $|q_4|^2$ is also periodic with $T_4 = 420 \pi^2$. 

The pattern of common multiple periods will generalize to the higher modes $q_n$ because of the rational relations between the interaction coefficients $C_{nmkl}$ in global AdS$_4$. Namely, one can apply the strategy of \cite{islands} to derive explicit, complicated formulas for $C_{nmkl}$ as functions of $n,m,k,l$. This leads, in particular, to the conclusion that
\begin{align}
&2\pi C_{0000} = 45,\\
&2\pi C_{0n0n} = \frac{108+90n-33n^2-58n^3-21n^4-2n^5}{n(n+1)(n+2)(n+3)} + 6n(2n+3)\left(\sum_{k=1}^{n}\frac{1}{k(2k-1)}+\frac{2}{2n+1}\right).\nonumber
\end{align}
Hence, all $C_{0n0n}$ are rationals divided by $\pi$, and therefore $q_n$, which oscillate with frequencies given by rational combinations of $C_{0n0n}$, have common periods of the form $\pi^2$ times a rational number. (As a direct consequence, $|q_n|^2$, which are what is important for our topic of energy returns, also have common periods of the same form, typically somewhat shorter than the common periods of $q_n$.)
The common period of $q_0, q_1,\ldots q_n$ grows rather rapidly with $n$ (for instance, we get estimates of order $10^{22}$ for $n=25$).

\begin{figure}[t!]
		\begin{subfigure}[b]{1\linewidth}
			\hspace{12mm}\includegraphics[scale=0.5]{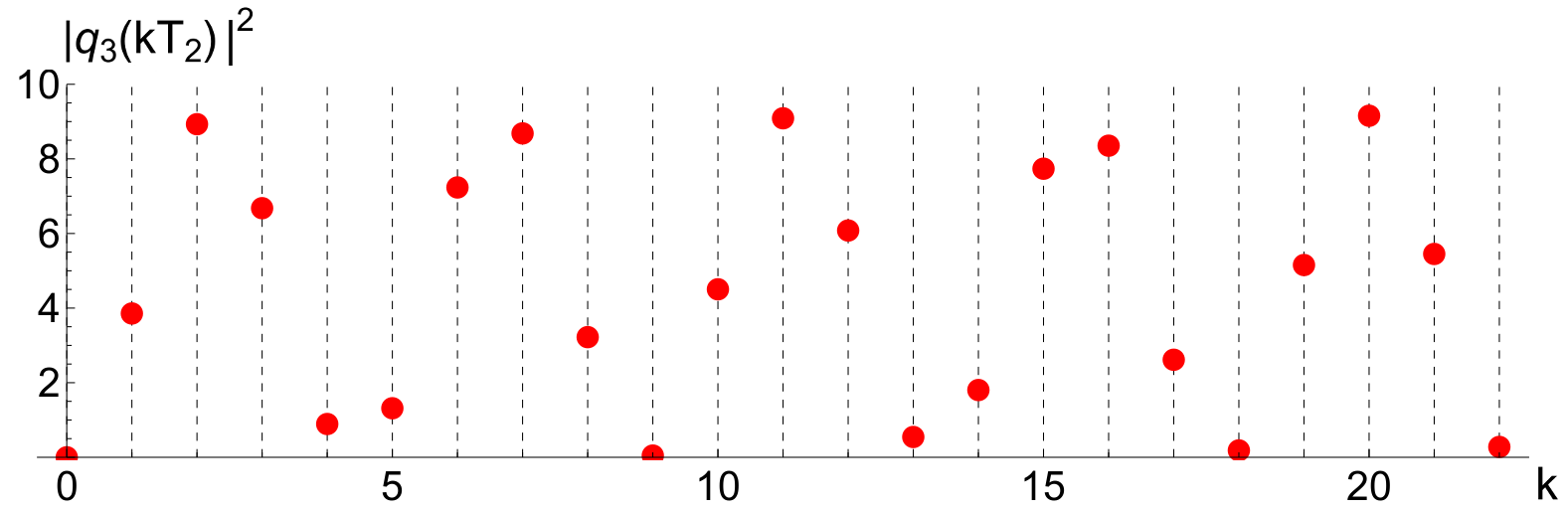}\vspace{-2mm}
			\caption{}
		\end{subfigure}\vspace{3mm}

		\begin{subfigure}[b]{.5\linewidth}
			\includegraphics[scale=0.28]{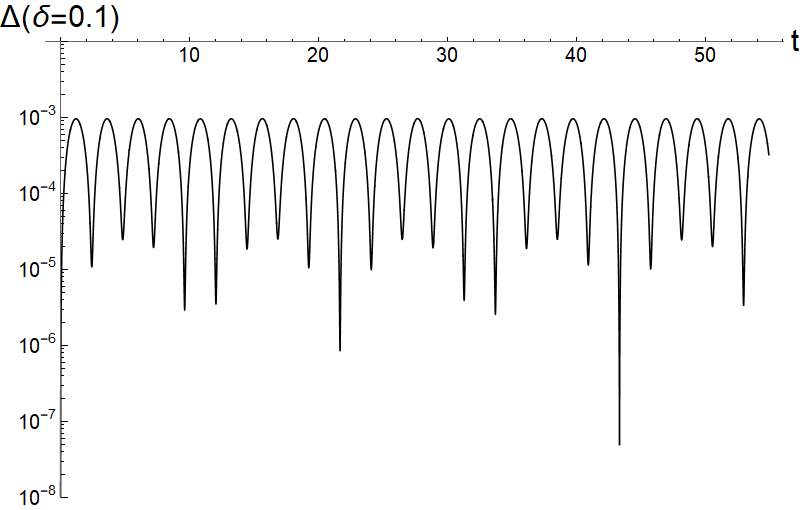}\vspace{-3mm}
			\caption{}
		\end{subfigure}  
		\begin{subfigure}[b]{.5\linewidth}
			\includegraphics[scale=0.28]{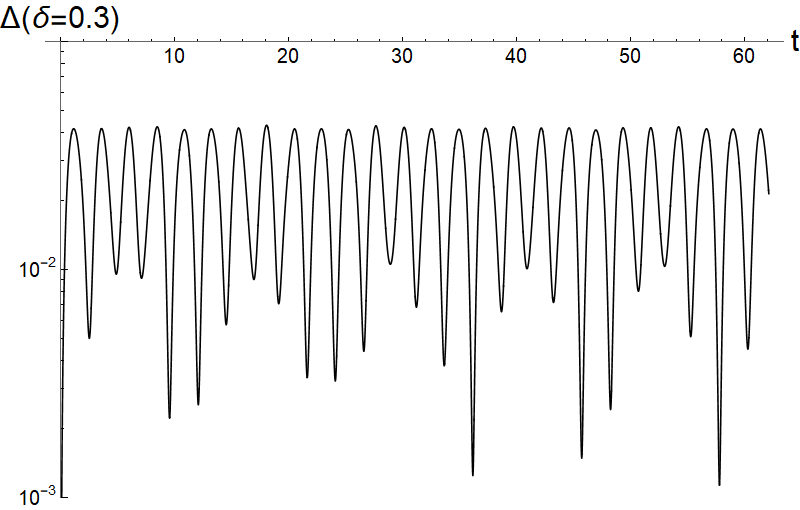}\vspace{-3mm}
			\caption{}\label{SLL0delta03}
		\end{subfigure}
		\caption{Analytics vs. numerics for two-mode initial data dominated by mode 0:\\ (a) analytic prediction for return accuracy based on (\ref{SLL0_q3}) and showing accurate returns after 4, 5, 13 and 14 oscillations, and extremely accurate returns after 9 and 18 oscillations;\\
(b) $\Delta$ defined by (\ref{deltadef}) for the numerical solution of the full resonant system (\ref{ressyst}) with two-mode initial data corresponding to $\de=0.1$ perfectly reproducing the analytic pattern;\\
 (c) the same for $\delta=0.3$, already quite far from the single-mode initial data limit, but still accurately reproducing returns after 4 and 5 oscillations.}
		\label{fig:SLL0}
	\end{figure}

There is a different form of accurate-but-imperfect returns, which is less spectacular than what we have described above in the limit $\de\to 0$, but more relevant for not-so-long times and not-so-small $\de$. As we demonstrated above, there are long multiple periods after which large sets of $q_{n\ge 2}$ simultaneously vanish, providing for very accurate energy returns in the lowest two modes. The basic `crude' return period is set by $T_2$, the period of $q_2$. At $t=kT_2$ with an integer $k$, $q_2=0$ and hence one gets approximate returns. If at this moment, the value of $q_3$ defined by (\ref{SLL0_q3}) is numerically small, even if it is not zero, one gets an improved return precision. In fig.~\ref{fig:SLL0}, we have performed comparisons of this analytic picture with numerical simulations of the resonant system (\ref{ressyst}). First, we have plotted $|q_3(kT_2)|^2$ and identified the specific small numbers of periods $k$ after which one expects improved return precision. We have then run numerical simulation of (\ref{ressyst}) with initial data $\alpha_0(0)=1/\sqrt{3(3+5\delta^2)}$,  $\alpha_1(0)=\de/\sqrt{5(3+5\delta^2)}$, which are a rescaled version of (\ref{SLL0defeq}) corresponding to the normalization (\ref{energynorm}). At $\de=0.1$, which corresponds to $1.6\%$ of the total energy initially in mode 1, the pattern of returns exactly matches our analytic predictions. At $\de=0.3$, which corresponds to $13\%$ of the total energy initially in mode 1, already very far from the strict $\de\to 0$ limit, the latter part of the return history is upset, but the first two accurate returns (after 4 and 5 oscillations) still match our analytic picture. Our treatment is thus robust, and retains predictive power even outside the region of very small $\de$.

	To summarize, the expansion of equation (\ref{ressyst}) in powers of $\delta$ around mode 0 enabled us to verify that the energy transfer between the modes is not exactly periodic for initial conditions close to mode 0. Nevertheless, there are rational relations between the AdS$_4$ interaction coefficients such that arbitrarily precise returns to the initial energy distribution starting from two-mode initial data\footnote{Note that we have not essentially used the assumption $q_{n\ge 2}=0$ in our reasoning, and hence the picture of arbitrarily precise returns over long periods should apply to more general initial data hierarchically suppressed away from mode 0 as in (\ref{SLL0defeq}).} sufficiently close to mode 0 occur if one waits long enough. Many other return patterns of varying accuracy are seen even for moderate initial ratios of energies of modes 0 and 1 in the initial state. This is very much in the spirit of the original FPU paradox. Corrections at subleading orders in $\de$ may be effectively considered starting with (\ref{resssystde}), and are likely to display many further patterns thanks to the ubiquitous presence of rational numbers in the problem, but we shall not explore this systematically.
	
	\subsection{Solutions dominated by  mode 1}\label{sec:SLL1}
	
	We shall now assume a hierarchically organized spectrum dominated by mode 1:
	\begin{equation}
	\alpha_{0} = \delta\, \frac{q_{0}(t)}{\sqrt{\om_0}}, \quad \alpha_{n \geq 1} = \delta^{n-1} \,\frac{q_{n}(t)}{\sqrt{\om_n}} ,
	\end{equation}
	with $\de\ll 1$.
	We then get from (\ref{ressyst}), after redefining $C_{nmkl}^{\rm new} = C_{nmkl}^{\rm old}/\sqrt{\om_n\om_m\om_k\om_l}$,  the following equation for mode 0:
	\begin{equation}
	i\dot{q}_{0} = \sum_{m=2}^{\infty}\sum_{k=1}^{m-1}C_{0mk,m-k}\bar{q}_{m}q_{k}q_{m-k}\delta^{2m-4} +2\sum_{m=1}^{\infty}C_{0m0m}\bar{q}_{m}q_{m}q_{0}\delta^{2m-2} + C_{0000}\bar{q}_{0}q_{0}q_{0}\delta^{2},
	\end{equation}
	and for all higher modes,
	\begin{equation}\nonumber
	i\dot{q}_{n} = \sum_{m=1}^{\infty}\sum_{k=1}^{n+m-1}C_{nmk,n+m-k}\bar{q}_{m}q_{n+m-k}q_{k}\delta^{2m-2} +\sum_{k=1}^{n-1}C_{n0k,n-k}\bar{q}_{0}q_{k}q_{n-k} + 2 \sum_{m=1}^{\infty}C_{0n0n}\bar{q}_{0}q_{0}q_{n}\delta^{2}. 
	\end{equation}
	Retaining only the leading terms, we obtain
	\begin{equation}
	i\dot{q}_{0} = C_{0211}\bar{q}_{2}q_{1}^2 +2C_{0101}|q_{1}|^{2}q_{0},
	\label{eq_SLL1_m0}
	\end{equation}
	and
	\begin{equation}
	i\dot{q}_{n} = \bar{q}_{1}\sum_{k=1}^{n}C_{n1k,n+1-k}q_{k}q_{n+1-k} +\bar{q}_{0}\sum_{k=1}^{n-1}C_{n0k,n-k}q_{k}q_{n-k}.
	\label{eq_SLL1_mn} 
	\end{equation}
	To appreciate the structure, we write out the first few equations explicitly:
	\begin{eqnarray}
	i \dot{q}_{0} & = & 2 C_{0101}|q_{1}|^{2}q_{0} + C_{0211}\bar{q}_2 q_{1}^2,\label{eq:SLL1_eq_0_exp} \\
	i \dot{q}_{1} & = & C_{1111}|q_{1}|^{2}q_{1}, \label{eq:SLL1_eq_1_exp}\\
	i \dot{q}_{2} & = & 2 C_{2121}|q_{1}|^{2}q_{2} + C_{2011}\bar{q}_0q_{1}^2,  \label{eq:SLL1_eq_2_exp}\\
	i \dot{q}_{3} & = & 2 C_{3131}|q_{1}|^{2}q_{3} + C_{3122}\bar{q}_1q_{2}^2   + 2C_{3021}\bar{q}_0 q_{1}q_{2}, \label{eq:SLL1_eq_3_exp}\\
	i \dot{q}_{4} & = & 2 C_{4141}|q_{1}|^{2}q_{4} + 2C_{4132}\bar{q}_1q_{2}q_{3}   + C_{4022}\bar{q}_0q_{2}^2. \label{eq:SLL1_eq_4_exp}
	\end{eqnarray}
	One first solves (\ref{eq:SLL1_eq_1_exp}), where one can set $|q_{1}|^{2} = 1$ as before using the scaling symmetry. After that, (\ref{eq:SLL1_eq_0_exp}) and (\ref{eq:SLL1_eq_2_exp}) form a system of two coupled linear equations for $q_0$ and $q_2$. Once $q_0$, $q_1$ and $q_2$ have been thus obtained, all the higher equations are solved recursively one-by-one, in a manner completely analogous to what we have previously described for solutions dominated by mode 0.
	
	Implementing this solution in practice for $q_0(0)=1$, $q_1(0)=1$ and $q_{n>1}(0) = 0$  (as before, the complex phases can be eliminated using the symmetries (\ref{symtransNE}), while the magnitude of $q_0(0)$ can be fixed by redefining $\de$), one gets
	\begin{eqnarray}
	q_{1}(t) & = & e^{-iC_{1111}t}, \\
	q_{0}(t) & = & \left(\cos\frac{\lambda t}{2} + \frac{ i \beta}{\lambda} \sin\frac{\lambda t}{2}\right) e^{-i(2C_{0101} + \beta/2)t},\\
	q_{2}(t) & = & \frac{-2i C_{1102}}{\lambda}\sin\frac{\lambda t}{2} e^{-i(2C_{2121} + \beta/2)t},
	\end{eqnarray}
	where
	$ \beta = 2\left(C_{1111}-C_{0101} - C_{2121}\right)$ and $\lambda = \sqrt{\beta^{2} - 4 C_{1102}^{2}}$. In what follows, we assume $\lambda\in \mathbb{R}$, which is the case for the system we are dealing with, as well as for all the systems studied in \cite{CF,BBCE,BEL,BBCE2,AO}.
	Note that, here, one gets nontrivial flows of energy among the first subleading modes, unlike the case dominated by mode 0. These flows of energy are always periodic with period $2\pi/\lambda$, as far as the three lowest modes are concerned.
	The first order at which violations of periodicity may enter is in the mode $q_3$, which is described by the following solution:
	\begin{equation}
	q_{3}(t) =  \frac{C_{1102}}{2\gamma\lambda^{2}(\gamma^2-\lambda^2)} \left(a + b e^{i \lambda t} + c e^{2i\lambda t} + d e^{i(\lambda + \gamma)t}\right)e^{i (-\lambda + 2C_{0101} - 2C_{2121} - C_{1111})t},
	\label{eq_SLL1_solution_n_3}
	\end{equation}
	with
	\begin{eqnarray}
	a & = &   2\gamma (\gamma - \lambda)\left(C_{1102} C_{2213} + (\beta +\lambda) C_{1203}\right), \label{eq:a_eq_SLL1_n4}\nonumber\\
	b & = &4(\lambda^2 - \gamma^2)\left( C_{1102}C_{2213} + \beta C_{1203}\right), \label{eq:b_eq_SLL1_n4}\nonumber\\
	c & = &  2\gamma (\gamma + \lambda)\left(C_{1102}C_{2213} + (\beta - \lambda) C_{1203}\right) ,\label{eq:c_eq_SLL1_n4}\\
	d & = & - 4 \lambda^{2}\left(C_{1102}C_{2213} + (\beta - \gamma) C_{1203} \right),  \label{eq:d_eq_SLL1_n4}\nonumber \\
	\gamma & = & -2C_{0101} + C_{1111} + 2C_{2121} - C_{3131}.\nonumber
	\label{eq:gamma_eq}
	\end{eqnarray}
	For the resonant system (\ref{ressyst}), direct computation yields $\lambda/\gamma = 7\sqrt{3665}/149 \approx 2.84$.  (Note that irrational numbers emerge here already in the periods of low-lying modes, unlike the case of initial data dominated by mode 0 we have considered previously.)

Suppose for a moment that one had $\lambda/\gamma = 3$.  Then $E_3\sim|q_3|^2$ computed from (\ref{eq_SLL1_solution_n_3}) would have been proportional to
	\begin{equation}
	\Big|a + b e^{i\lambda t} + c e^{2i\lambda t} + d e^{4i \lambda t/3}\Big|^2,
	\end{equation}
	which oscillates with period of $6\pi/\lambda$, which is thrice the period of $|q_0|^2$,  $|q_1|^2$, $|q_2|^2$. One would thus have found exact returns after three direct-reverse cascades as far as the first four modes are concerned.

	Now, for the actual resonant system we study, $\lambda/\gamma$ is of course not exactly 3. This shows that the third energy return cannot be exact, at least for initial data close to mode 1, no matter how close we get to exact returns in our numerical simulations. At the same time, the fact that 2.84... is close to 3 explains why we are seeing very accurate returns after three direct-reverse cascades (and also why the returns after the first two direct-reverse cascades are less exact), as in figs.~\ref{fig:AdS_9_oscillations} and \ref{fig:two_mode_evol_sequence_D}.
	\begin{figure}[t!]
		\begin{subfigure}[b]{1\linewidth}
			\hspace{8mm}\includegraphics[scale=0.5]{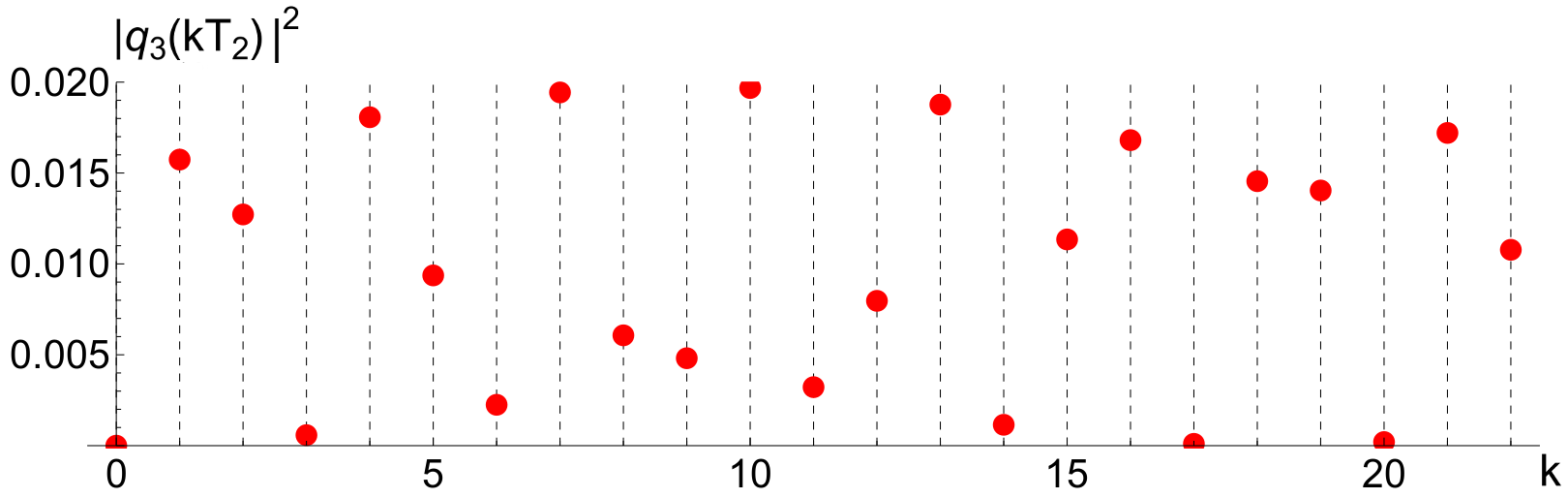}\vspace{-2mm}
			\caption{}
		\end{subfigure}\vspace{3mm}

		\begin{subfigure}[b]{.5\linewidth}
			\includegraphics[scale=0.28]{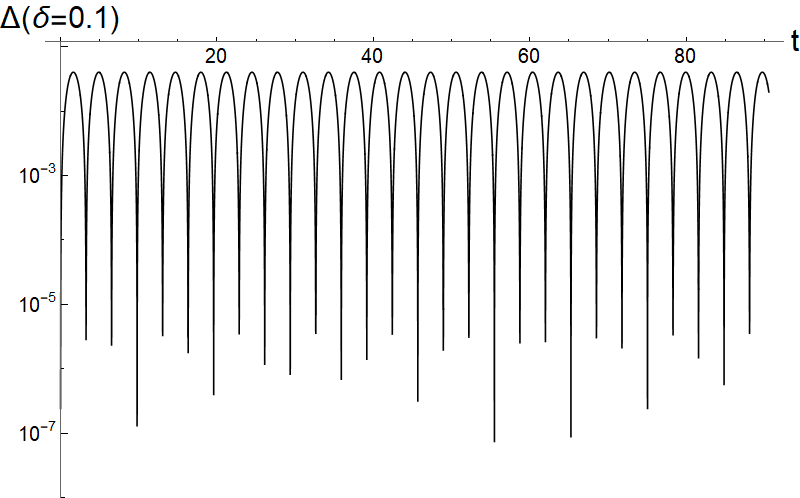}\vspace{-3mm}
			\caption{}
		\end{subfigure}  
		\begin{subfigure}[b]{.5\linewidth}
			\includegraphics[scale=0.28]{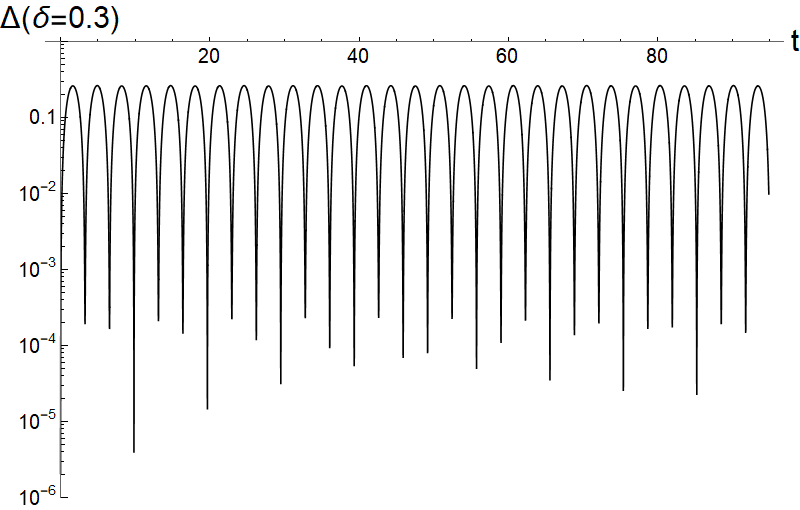}\vspace{-3mm}
			\caption{}
		\end{subfigure}
		\caption{Analytics vs. numerics for two-mode initial data dominated by mode 1:\\ (a) analytic prediction for return accuracy based on (\ref{eq_SLL1_solution_n_3}) and showing accurate returns after 3, 6 and 14 oscillations, and extremely accurate late-time returns after 17 and 20 oscillations;\\
(b) $\Delta$ defined by (\ref{deltadef}) for the numerical solution of the full resonant system (\ref{ressyst}) with two-mode initial data corresponding to $\de=0.1$ perfectly reproducing the analytic pattern;\\
 (c) the same for $\delta=0.3$, already quite far from the single-mode initial data limit, but still accurately reproducing returns after 3 and 6 oscillations.}
		\label{fig:SLL1}
	\end{figure}

One can extend the above argument into a more quantitative analysis by plotting $|q_3(kT_2)|^2$ at integer $k$ with $q_3$ given by (\ref{eq_SLL1_solution_n_3}) and $T_2=2\pi/\lambda$, as we did before for solutions dominated by mode 0. Since $q_2(kT_2)=0$ by construction, having a small $q_3$ at the same moment signifies  a return of enhanced  precision. We have displayed the results of this analysis in fig.~\ref{fig:SLL1}. As in the previous section, at $\de=0.1$, our predictions are perfect, and at $\de=0.3$, already quite far from the single-mode data limit, we still accurately predict the first two returns.
	
	Note that, in general, our analysis gives a nice perspective on why direct-reverse cascade oscillations are ubiquitously seen in numerical simulations of various resonant systems. For solutions dominated by mode 1, for example, the first three modes perform an infinite sequence of direct-reverse cascades, while the higher modes hold only a small amount of energy and provide cosmetic modifications to the cascades. (Furthermore, even these higher modes oscillate with frequencies comparable to the lowest modes.)

	\section{Discussion}\label{sec:discussion}
	
	We have revisited the issue of energy returns to the initial configuration for two-mode initial data in the resonant approximation to weakly nonlinear gravitational dynamics of the AdS$_4$-scalar-field system. Having performed numerical simulations with much higher precision than what has been previously seen in the literature, we have observed returns of striking accuracies, exemplified by fig.~\ref{fig3periods}. The numerics also provided indications, however, that the small imperfections we see cannot be purely due to numerical artifacts. To elucidate the situation, we have performed an analytic study of solutions dominated by one of the two lowest modes and proved that the accurate returns we have observed numerically are inexact in this limit dominated by one of the two modes. This should be contrasted with a scenario observed in the recent literature on related systems \cite{CF,BBCE,BEL,BBCE2,AO}, where perfect returns occur for all two-mode initial data of the form~(\ref{twomode}).
	
\begin{figure}[t]
		\centering
		\includegraphics[scale=0.6]{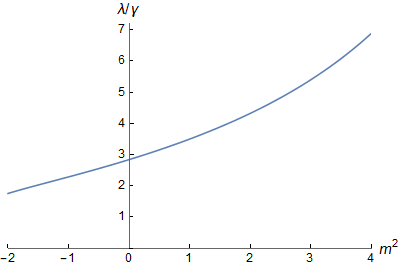}  
		\caption{Dependence of $\lambda/\gamma$ on the mass of the scalar field. Adjusting $m^2$, $|q_3|^2$ can be made exactly periodic. For example, on can choose $m^2$ such that $\lambda/\gamma = 3$, as briefly considered in section \ref{sec:SLL1}.}
		\label{fig:lambdaovegamma}
	\end{figure}

	As usual in FPU-like situations, it is natural to expect that the near-perfect returns arise due to proximity to another dynamical system for which the returns are exact. One could first try to improve the quality of returns by variation of parameters of the model, such as the dimension of AdS or the mass of the scalar field. We have explored this scenario in the context of our analytic treatment of solutions dominated by mode 1 in section~\ref{sec:SLL1}. We have seen that by adjusting the mass of the scalar field, it is possible to make the return of mode 3 exact (fig.~\ref{fig:lambdaovegamma}). This evidently qualitatively improves the precision of returns for initial data near mode 1, as the discrepancy is now in the strongly suppressed modes starting from mode 4, but the scenario still falls short of providing exact returns. We also note that a nonrelativistic version of the AdS dynamics (which technically corresponds to the limit of infinite scalar field mass) can be analytically proved to display perfect returns \cite{SNH} in AdS$_5$, rather than in AdS$_4$ (returns in AdS$_5$ at finite masses, on the other hand, are not close to being perfect).
	More broadly, one could look for arbitrary small modifications of the interaction coefficients $C$ in (\ref{ressyst}), irrespectively of whether they originate from standard physically motivated PDEs, and perfect returns in any such system would be sufficient to clarify the origin of near-perfect returns in the (physically motivated) resonant system (\ref{ressyst}). In \cite{AO}, a very large class of resonant systems displaying perfect returns for two-mode initial data has been constructed. This class is characterized, in particular, by the following identities satisfied by the interaction coefficients:
	\begin{align}
	\sum_{k=0}^{n+m}\frac{f_kf_{n+m-k}}{f_nf_m}C_{nmk,n+m-k}& =1,\label{idpoly}\\
	\sum_{k=0}^{n+m}k^2\frac{f_kf_{n+m-k}}{f_nf_m}C_{nmk,n+m-k}& = c_2(n^2+m^2) + c_1 n m + c_0 (n+m),\nonumber
	\end{align}
	where $f_n$ can be either $1/\sqrt{n!}$ or $\sqrt{G(G+1)\cdots(G+n-1)/n!}$, and $G$, $c_0$, $c_1$, $c_2$ are arbitrary constants. Our numerical comparisons indicate that the interaction coefficients of the resonant system (\ref{ressyst}) do not appear close to satisfying such relations. One could look for modifications of (\ref{idpoly}) approximately satisfied by the AdS$_4$ interaction coefficients. Our preliminary study shows that some polynomial-type summation identities are satisfied with a good precision, but we feel that it is premature to judge what exact dynamical implications follow from such approximate identities.

We comment explicitly on what we have achieved in comparison to \cite{FPU}, where the topic of FPU-like behaviors in weakly nonlinear AdS$_4$ dynamics was first brought up. In \cite{FPU}, the focus was on reporting, among other things, the return after three direct-reverse cascades for two-mode initial data with equal energies that can be seen in our fig.~\ref{fig:two_mode_evol_sequence_D}. Since the simulations were performed with a rather limited precision, the results were in principle consistent with perfect returns upset exclusively by numerical artifacts, especially in light of the subsequent discovery of closely related systems for which the returns are exact. We have now ruled out this possibility. At the same time, we have observed imperfect returns of precision much higher than what is suggested by the material of \cite{FPU} and repeating over many oscillation cycles, as seen in fig.~\ref{fig3periods}. More importantly, our analytic investigations of section 4 have generated a neat picture of returning behaviors for initial data sufficiently close to mode 0 or mode 1, and allowed for identification of specific multiple oscillation periods after which enhanced returns occur. This picture remains valid even for initial data with moderate energy ratios of the two lowest modes. The rational relations between the AdS$_4$ interaction coefficients we have displayed allude to hierarchies of returns, with specific longer waiting times resulting in returns of better and better accuracy. One in fact sees returns of striking accuracy at late times in fig.~\ref{SLL0delta03}, outside the domain where our concrete analytic approximation are valid. This suggests that there are further structures to be explored. Elaborate patterns of returns of varying precision over long times are known from the original FPU problem. (The FPU chain is now believed to thermalize over very long times \cite{FPUtherm}, but in our resonant system (\ref{ressyst}), the recurrences are likely to persist forever.)
	
	The AdS/CFT paradigm provides an intriguing potential link between our results and dynamics of conformal field theories (CFTs) on spatial spheres. In particular, one would expect that very close returns to the initial state must occur at low energies in CFTs that can be accurately approximated in an appropriate `holographic' limit by a single scalar field in the AdS bulk coupled to gravity. It would be very interesting to delineate this class of systems more precisely, and build explicit connections to the sort of dynamics we have described in this article. There is recent literature \cite{FHHPRV,OPS} providing detailed links between evolution of CFT states and weakly nonlinear gravitational dynamics in the bulk, which is likely to be useful in this regard.
	
	\section*{Acknowledgments}
	
	We thank Piotr Bizo\'n and Javier Mas for discussions and for collaboration on related subjects. This research has been supported by  CUniverse research promotion project (CUAASC), by FWO-Vlaanderen (projects G044016N and G006918N), by Vrije Universiteit Brussel through the Strategic Research Program ``High-Energy Physics,'' by FPA2014-52218-P~from Ministerio de Economia y Competitividad, by Xunta de Galicia ED431C 2017/07, by the European Regional Development Fund (FEDER) and by Grant Mar\'\i a de Maeztu Unit of Excellence MDM-2016-0692. This research
	has benefited from the use computational resources/services provided by the Galician Supercomputing Centre (CESGA). A.B. thanks the Spanish program ``ayudas para contratos predoctorales para la formaci\'on de doctores 2015'' and its mobility program for his stay at Vrije Universiteit Brussel, where part of this project was developed.
	
\appendix	
	\section*{Appendix: Interaction coefficients for a massive scalar field in $\text{AdS}_{d+1}$}\label{Appenix:A}
	
	We shall consider a massive scalar field in $AdS_{d+1}$ with Dirichlet boundary conditions $\phi(\pi/2) = 0$, which made an appearance in section~\ref{sec:discussion}. This model is also fully resonant and an approximation of the form (\ref{ressyst}) can be derived. The massless case was developed in \cite{CEV1}; the process is quite similar for the massive scalar, so we will only present the main results. 
	
	The model is governed by the action
	\begin{equation}
	S = \frac{1}{16\pi G} \int d^{d+1}x \sqrt{-g}\left(R-2\Lambda\right) - \frac{1}{2}\int d^{d+1}x \sqrt{-g} \left(\partial_\mu\phi \partial^{\mu}\phi + m^2 \phi^2\right).
	\end{equation}
 Our ansatz for the metric is (\ref{eqn:MetricAnsatz}), 
with the same conventions as in section \ref{sec:preliminaries}.
Developing the time-averaging strategy described in the main text, or in \cite{CEV1,CEV2} which contain a more thorough discussion, we obtain the following expressions for the interaction coefficients, after splitting them in three types according to the number of coincident indices, $T_{l} \equiv C_{llll}$, $R_{il} \equiv 2C_{ilil}$ for $i\neq l$ and $S_{ijkl} \equiv C_{ijkl}$ for $\{i,j\}\neq \{k,l\}$:
	\begin{align}
	T_l =& \frac{1}{2}\om_l^2X_{llll}+\frac{3}{2}Y_{llll}+2\om_l^4W_{llll}+2\om_l^2W_{llll}^{*}-\om_l^2 \left(A_{ll} + \om_l^2V_{ll}\right),\\
	R_{il} =& \frac{1}{2}\left(\frac{\om_i^2+\om_l^2}{\om_l^2-\om_i^2}\right)\left(\om_l^2X_{illi} - \om_i^2 X_{liil}\right) + 2\left(\frac{\om_l^2 Y_{ilil} - \om_i^2 Y_{lili}}{\om_l^2-\om_i^2}\right) + \left(\frac{\om_i^2\om_l^2}{\om_l^2-\om_i^2}\right)\left(X_{illi}-X_{lili}\right)\nonumber\\
	& + \frac{1}{2}\left(Y_{iill}+Y_{llii}\right) + \om_i^2\om_l^2 \left(W_{llii}+W_{iill}\right) + \om_i^2 W_{llii}^{*} + \om_l^2 W_{iill}^{*} - \om_l^{2} \left(A_{ii} + \om_i^2 V_{ii}\right),\\
	S_{ijkl} =& -\frac{1}{4}\left(\frac{1}{\om_i+\om_j}+\frac{1}{\om_i - \om_k} + \frac{1}{\om_j-\om_k}\right)\left(\om_i\om_j\om_kX_{lijk} - \om_l Y_{iljk}\right)\nonumber\\ 
	&-\frac{1}{4}\left(\frac{1}{\om_i+\om_j}+\frac{1}{\om_i - \om_k} - \frac{1}{\om_j-\om_k}\right)\left(\om_j\om_k\om_lX_{ijkl} - \om_i Y_{jikl}\right)\nonumber\\
	& - \frac{1}{4}\left(\frac{1}{\om_i+\om_j}-\frac{1}{\om_i - \om_k} + \frac{1}{\om_j-\om_k}\right)\left(\om_i\om_k\om_lX_{jikl} - \om_j Y_{ijkl}\right)\nonumber\\
	& -\frac{1}{4}\left(\frac{1}{\om_i+\om_j}-\frac{1}{\om_i - \om_k} - \frac{1}{\om_j-\om_k}\right)\left(\om_i\om_j\om_lX_{kijl} - \om_k Y_{ikjl}\right),
	\end{align}
	where 
	\begin{align}
	&X_{ijkl} = \int_{0}^{\pi/2}dx e_{i}'(x)e_j(x)e_k(x)e_l(x)\mu(x)^2\nu(x),\\
	&Y_{ijkl} = \int_{0}^{\pi/2}dx e_{i}'(x)e_j(x)e_k'(x)e_l'(x)\mu(x)^2\nu(x),\\
	&W_{ijkl} = \int_{0}^{\pi/2}dx e_{i}(x)e_{j}(x)\mu(x)\nu(x)\int_{0}^{x}dy e_k(y)e_l(y)\mu(y),\\
	&W_{ijkl}^{*} = \int_{0}^{\pi/2}dx e_{i}'(x)e_{j}'(x)\mu(x)\nu(x)\int_{0}^{x}dy e_k(y)e_l(y)\mu(y),\\
	&V_{ij} = \int_{0}^{\pi/2} dx e_i(x)e_j(x)\mu(x)\nu(x),\\
	&A_{ij} = \int_{0}^{\pi/2} dx e_i'(x)e_j'(x)\mu(x)\nu(x).
	\end{align}
	Here, $e_n(x)$  and $\om_n$ are the eigenmodes and their associated eigenvalues of the linear problem:
	 \begin{equation}
	 \ddot{\phi}_1 + \hat{L} \phi_1 = 0 \qquad \text{with} \qquad  \hat{L} = -\frac{1}{\mu(x)}\partial_{x}\left(\mu(x) \partial_x\right)+\frac{m^{2}}{\cos^{2}{x}}.
	 \label{eq:linear_eq_mass}
	 \end{equation}
	 Their expressions are
	 	\begin{equation}
	 	e_{n}(x)=k_n\cos^{\,\Delta}x\,P_{n}^{\left(\frac{d}{2}-1,\ \Delta-\frac{d}{2}\right)}(\cos{2x}), \qquad k_n = 2\sqrt{\frac{(n+\Delta/2)\Gamma(n+1)\Gamma(n+\Delta)}{\Gamma(n+d/2)\Gamma(n+\Delta-d/2+1)}},
	 	\label{eq:eigenfunctions_mass}
	 	\end{equation}
	 	and 
	 		\begin{equation}
	 		\om_{n}=  \Delta + 2n, \qquad n=0,1,\ldots,
	 		\label{eq:eiguenvalues_massive_modes}
	 		\end{equation}
	 	where $\Delta$ satisfies the equation $\Delta(\Delta-d) = m^{2}$ and $P_{n}^{(a,b)}(x)$ are Jacobi polynomials. We have defined $\mu(x)$ and $\nu(x)$ as in (\ref{munu}).	 	

In analogy with the explanation for the massless scalar field given in section \ref{sec:preliminaries}, the current model has the resonant condition $\om_n+\om_m=\om_k+\om_l$, which through (\ref{eq:eiguenvalues_massive_modes}) is equivalent to $n+m=k+l$. However there could be two more resonant channels, $\om_n=\om_m+\om_k+\om_l$ and $\om_n + \om_m + \om_k = \om_l$, equivalent to $n = m+k+l + \Delta$ and $n = l - m - k - \Delta$, respectively. If $\Delta$ is integer these last two conditions can be satisfied and a priori two new terms must be included in the system of equations  (\ref{ressyst}):
	\beq
	i\omega_n\dot\alpha_n=\hspace{-5mm}\sum_{\om_n+\om_m=\om_k+\om_l}\hspace{-7mm} C_{nmkl}\,\bar\alpha_m\alpha_k\alpha_l +\hspace{-5mm}\sum_{\om_n=\om_m + \om_k+\om_l}\hspace{-7mm} Q_{nmkl}\, \alpha_m\alpha_k\alpha_l +\hspace{-5mm} \sum_{\om_n+\om_m + \om_k=\om_l}\hspace{-7mm} U_{nmkl}\,\bar\alpha_m\bar\alpha_k\alpha_l.
	\label{ressyst_full}
	\eeq	
	In our situation we did not perform an analytic study of $Q_{ijkl}$ and $U_{ijkl}$ as in \cite{CEV2}, where it was proven that  for a massless scalar field ($\Delta = d$) these coefficients vanish, but numerical calculations suggest that they also vanish for nonzero masses. On the other hand, when $\Delta$ is not integer, the conditions $n = m+k+l + \Delta$ and $n = l - m - k - \Delta$ are not satisfied for any combination of the indices. Therefore, these interaction channels disappear upon time-averaging. We thus see that, for any $\Delta$, the  relevant dynamics is governed by equation (\ref{ressyst}) through $T_l$, $R_{il}$ and $S_{ijkl}$.
	
	We note that the given expressions for $T_l$, $R_{il}$ and $S_{ijkl}$ are exactly the same as in \cite{CEV2}, where all derivations are specialized to $m^2=0$. However, if we were to transform these integrals using integration by parts to the form of \cite{CEV1}, the resulting expressions would have differed from those of \cite{CEV1} by additional terms with explicit dependence on $m^2$.
	

\end{document}